\documentclass[lettersize,journal]{IEEEtran}
\usepackage{amsmath,amsfonts}

\usepackage[linesnumbered,ruled]{algorithm2e}

\usepackage{array}
\usepackage[caption=false,font=small,labelfont=rm,textfont=rm]{subfig}
\usepackage{textcomp}
\usepackage{stfloats}
\usepackage{url}
\usepackage{verbatim}
\usepackage{graphicx}
\usepackage{cite}
\usepackage{multirow}
\usepackage{booktabs}
\usepackage{makecell}
\usepackage{amssymb} 
\usepackage{cases} 
\usepackage{hyperref}  
\usepackage{tabularx}
\hypersetup{hypertex=true,
	colorlinks=true,
	linkcolor=blue,
	anchorcolor=blue,
	citecolor=blue}
\usepackage{booktabs}
\usepackage[table]{xcolor}

\begin{document}

\title{Multi-Target Flexible Angular Emulation for \\ ISAC Base Station Testing Using a \\  Conductive  Amplitude and Phase Matrix Setup: Framework and Experimental Validation }

\author{Chunhui Li, Chengrui Wang, Zhiqiang Yuan, and  Wei Fan
\thanks{    This work was supported by the National Natural Science Foundation of
	China under Grant  62571119. \textit{(Corresponding author: Wei Fan.)}}
\thanks{Chunhui Li and Chengrui Wang are with the National Mobile Communications Research Laboratory,	School of Information Science and Engineering, Southeast University, Nanjing 210096, China (e-mail: lichunhui@seu.edu.cn; 220251014@seu.edu.cn). } 

\thanks{ Zhiqiang Yuan and Wei Fan are with the National Mobile Communications Research Laboratory,	School of Information Science and Engineering, Southeast University, Nanjing 210096, China, and also with the Purple Mountain Laboratories,  Nanjing 211111, China (e-mail: zqyuan@seu.edu.cn; weifan@seu.edu.cn). }
}



\maketitle

\begin{abstract}

Comprehensive evaluation of the functionalities, algorithms, hardware components, and performance characteristics of future integrated sensing and communication (ISAC) base stations (BSs) under realistic deployment scenarios in controlled laboratory environments represents a critical requirement for ISAC technology advancement. A primary challenge in achieving this objective involves the emulation of multiple targets with arbitrary radar cross-section (RCS), range, angle, and Doppler profiles for ISAC BS equipped with large-scale antenna arrays using radar target simulator (RTS) with limited interface ports.
In this work, we introduce a simple yet highly effective and practical conductive amplitude and phase matrix framework to address this fundamental challenge. 
The core concept involves introducing a tunable conductive amplitude and phase modulation network in the test configuration between the ISAC BS under test  and a RTS. Based on this structure, we subsequently investigate the corresponding configurations for different sensing operational modes of ISAC BSs, specifically the array duplex transmission and reception (ADTR) mode and the split-array transmission and reception (SATR) mode.
For experimental validation, we design two distinct monostatic sensing scenarios to demonstrate the framework capabilities across both operational modes. The first scenario involves dynamic multi-drone sensing validation for ADTR mode operation, while the second scenario addresses static single-drone sensing for SATR mode validation. The experimental results demonstrate that the proposed framework can accurately emulate the joint RCS, range, velocity, and angular characteristics of multiple sensing targets within the conductive test environment, highlighting its significant potential for testing applications in sub-6 GHz ISAC BS development and validation.

\end{abstract}

\begin{IEEEkeywords}
	Integrated sensing and communication (ISAC), radar target emulation, ISAC BS testing, 6G
\end{IEEEkeywords}

\section{Introduction}
\IEEEPARstart{I}{n} recent years, integrated sensing and communication (ISAC) has become a key  candidate technology for sixth-generation (6G) communication systems, enabling the deep integration of communication and sensing \cite{wang2023road}. The ISAC base station (BS)  is expected  to accurately estimate the range, velocity, direction, and radar cross-section (RCS) of sensing targets to support applications such as low-altitude traffic surveillance, maritime monitoring, navigation management, and the metaverse \cite{dong2022sensing,liu2022integrated,Chen2024}. These capabilities, which are absent in conventional communication BSs, present a new challenge for the research community, namely,  the comprehensive evaluation of  ISAC BSs to determine their actual sensing performance under deployment scenarios in the laboratory \cite{wang2025channel}.

One commonly adopted test approach is field testing, which uses real targets or equivalent physical models in practical environments to evaluate the sensing capability of ISAC BSs \cite{zhang2022time, yang2024isac}. 
While this method is fundamental and necessary, it suffers from inherent limitations
\cite{koopman2016challenges}.
First, field testing is time- and labor-intensive, as ISAC BSs and sensing targets must be deployed and adjusted at designated test sites, often under non-laboratory conditions. 
Second, repeatability is limited, particularly in scenarios with multiple moving targets, as it is difficult to guarantee identical test conditions  across experiments. 
Lastly, the high cost of constructing realistic  test environments restricts the number of feasible test cases.

Another technique is to emulate virtual sensing targets to deceive the device under test (DUT) \cite{gadringer2018virtual}. A key instrument for this approach is the radar target simulator (RTS), which receives signals through its RF front ends and, via its internal signal processing system, applies delay, Doppler shift, and attenuation before re-radiating the modified signals through the front ends.
Recent studies in the radar community have investigated RTSs with diverse functionalities. For accurate moving target simulation, Körner \textit{et al.} \cite{korner2021multirate} proposed a multirate universal RTS, while for wideband radar emulation, Zhai \textit{et al.} \cite{zhai2024wideband} presented a wideband photonic RTS based on an all-optical IQ upconverter.
Additionally, the authors in \cite{wang2025channel} propose a cost-effective testing framework that repurposes a channel emulator (CE) to replace the RTS for generating deceptive echoes.
  Concurrently, extensive efforts have been devoted to emulating the angular characteristics of radar targets. Buddappagari \textit{et al.}  \cite{buddappagari2021over} and Asghar \textit{et al.}  \cite{asghar2021radar} employed target simulators where RF front-ends are mechanically repositioned to achieve the intended angle-of-arrival (AoA). However, this approach imposes stringent requirements on the mechanical structure, constrains the allowable target motion rate. Scheiblhofer \textit{et al.}  \cite{scheiblhofer2017low} and Gadringer \textit{et al.}  \cite{gadringer2018radar} utilized electronically switched front-ends for radar target simulation without mechanical movement. Nevertheless, this method necessitates a substantial number of front-ends to ensure adequate angular resolution.
  To address the limitations of the aforementioned approaches and achieve flexible radar target spatial characteristic emulation, Diewald \textit{et al.}  \cite{diewald2021arbitrary} proposed an RTS supporting arbitrary AoA simulation based on the field synthesis concept. Furthermore, Schroeder \textit{et al.}  \cite{schoeder2021flexible,schoeder2023unified} presented an over-the-air flexible direction-of-arrival RTS that performs calibration by applying the inverse propagation matrix relating the RTS  front-ends to the DUT antenna array.

  \begin{table*}[!t]
 	\centering
 	\caption{Comparison of Related RTS Approaches and This Work}
 	\label{tab:radar-emulation-comparison}
 	\footnotesize 
 	\setlength{\tabcolsep}{2.2pt} 
 	\renewcommand{\arraystretch}{1.2} 
	\begin{tabularx}{\textwidth}{@{}lcccrrp{4.5cm}@{}}
 		\toprule
 		\textbf{Reference} & \textbf{Category} & \textbf{Angular Emulation  Flexibility} & \textbf{Delay \& Doppler Emulation} & \textbf{DUT Configuration} & \textbf{Test Frequency} \\
 		\midrule
 		\cite{buddappagari2021over}, \cite{asghar2021radar} & OTA (Mechanical motion) & Limited & Yes & ARS 510 (3Tx, 4Rx) & 77, 79 GHz \\
 		\addlinespace[0.5em]
 		\cite{scheiblhofer2017low} & OTA (Electronic switching) & Limited & Yes  & ADF5904/1 (1Tx, 4Rx)  & 24 GHz\\
 		\addlinespace[0.5em]
 		\cite{gadringer2018radar} & OTA (Electronic switching) & Limited & Yes  &  ARS408 (2Tx, 6Rx) & 77 GHz  \\
 		
 		\addlinespace[0.5em]
 		\cite{diewald2021arbitrary} & OTA (Field synthesis)   & High & Yes  & AWR1843 (2Tx, 4Rx) & 77 GHz \\
 		\addlinespace[0.5em]
 		\cite{schoeder2021flexible,schoeder2023unified} & OTA (Propagation matrix inversion) & High & Yes  & \raggedright 1Tx, 4Rx\cite{schoeder2021flexible}, 6Rx \cite{schoeder2023unified} & 78 GHz \\
 		\addlinespace[0.5em]
 		\cite{wang2025channel} & OTA/ Conducted & N/A & Yes  & 1 Tx, 1 Rx & 3.5 GHz \\
 		\addlinespace[0.5em]
 		\textbf{This work} & \textbf{Conducted} & \textbf{High}  & \textbf{Yes} & \textbf{32 Tx, 32 Rx }& \textbf{3.5 GHz} \\
 		\bottomrule
 	\end{tabularx}
 \end{table*}

 However, existing RTS schemes with flexible AoA emulation from the radar community encounter several  challenges when applied to ISAC BSs. 
 Currently, most RTS solutions primarily address automotive millimeter-wave radars operating in the V-band with relatively few antenna elements. In contrast, current ISAC BSs operate at lower frequencies, predominantly in sub-6 GHz bands, exhibit larger physical apertures, and employ large-scale antenna arrays to achieve high angular resolution capabilities.
For existing field synthesis-based RTS methods \cite{diewald2021arbitrary}, achieving high angular resolution necessitates densely populated probe arrays, while for propagation matrix inversion-based RTS approaches \cite{schoeder2021flexible,schoeder2023unified}, the large number of DUT antennas causes the propagation matrix condition number to increase rapidly, severely compromising emulation accuracy \cite{zhang2020achieving}.
 The primary limiting factor for existing schemes stems from the integrated design of automotive millimeter-wave radar systems, which necessitates OTA testing without RF cable connections, whereas existing sub-6 GHz ISAC BSs typically provide antenna connectors enabling conductive RF cable connections between test equipments and the DUT ISAC BS.
 Additionally, in our previous work \cite{wang2025channel}, which demonstrated that CE previously employed for communication testing can be reconfigured for sensing test and established its capabilities and limitations in the Doppler, radar cross-section (RCS), and range domains, the spatial domain remained unexplored.
 In this paper, we introduce a conductive testing framework to realize multi-target flexible angular emulation using a tunable amplitude and phase matrix, also referred to as an amplitude and phase modulation (APM) network, for sub-6 GHz ISAC BSs equipped with large-scale antenna arrays. A comparative summary of the proposed framework with existing relevant RTS schemes is presented in Table \ref{tab:radar-emulation-comparison}.
 
The main contributions of this work are listed as follows:
\begin{itemize}
	\item First, we propose a simple yet highly effective and practical conductive testing framework for multi-target flexible angular emulation in ISAC BS  under test. The core concept introduces an APM network in the test setup between the DUT ISAC BS and a RTS. This approach enables the emulation of multiple targets with arbitrary RCS, range, angle, and Doppler profiles for ISAC BSs equipped with large-scale antenna arrays using RTSs with limited interface ports.
	\item Additionally, considering that ISAC BSs are anticipated to support distinct sensing modes, namely array duplex transmission and reception (ADTR) mode and split-array transmission and reception (SATR) mode, we investigate the corresponding configurations of the proposed framework for each sensing mode. For each configuration, corresponding signal models and tailored target emulation methodologies are developed.
	\item Finally, two representative sensing scenarios are designed to validate the respective configurations, wherein the sensing targets are experimentally emulated using the proposed framework and subsequently estimated by the emulated ISAC BS under test. Experimental results verify the feasibility of the proposed framework for multiple sensing target emulation in sub-6 GHz ISAC BS testing.
\end{itemize}

The remainder of this article is organized as follows. Section II presents the problem formulation, introduces the proposed framework with its distinct configurations, and establishes the corresponding signal models. Section III describes the experimental design and validation of the proposed framework. Finally, Section IV concludes the article.

\textit{Notations}:
Bold uppercase characters $\mathbf{X}$  denote matrices;
bold lowercase characters $\mathbf{x}$ denote vectors;
 $(\cdot )^{{T}}$ is the transpose operator; $t$ and $f$ denote the time and frequency, respectively.

\section{Framework}
\subsection{Problem Statement}

Our objective is to emulate multiple radar sensing targets for the ISAC BS under test, characterized by distinct range, RCS, velocity, and AoA parameters. This capability enables comprehensive performance evaluation and optimization of ISAC BSs under diverse operational scenarios.
The primary challenge lies in accurately emulating the angular domain characteristics of multiple targets. Current commercial RTSs cannot directly address this requirement \cite{rohde_schwarz_2024_areg800a}, as these instruments inherently focus on range, RCS, and velocity parameters, while angular domain emulation necessitates coordinated RF front-end configurations.
 Furthermore, as previously discussed, ISAC BSs typically employ large-scale antenna arrays, rendering existing OTA flexible AoA emulation  RTS methods from the radar community impractical.
Fortunately, sub-6 GHz ISAC BSs provide accessible conductive test interfaces between the RF chains and antenna elements, which constitutes the fundamental advantage leveraged by the approach presented in this work.

\begin{figure}[!t]
	\centering
	{\includegraphics[width=0.49\textwidth]{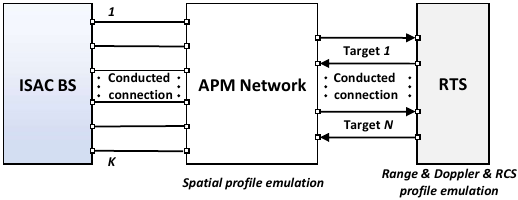}}
	\caption{Diagram of  the proposed sensing targets emulation framework for ISAC BS conducted testing, where $K$ represents the number of antenna ports and $N$ denotes the number of emulated targets. Note that for typical ISAC BS testing scenarios, the condition $K \gg N$ holds. }
	\label{fig_framework} 
\end{figure}
\subsection{The Proposed Framework}
As shown in Fig. \ref{fig_framework}, the proposed multiple sensing target emulation framework consists of three components: the  ISAC BS under test, the APM network, and the RTS. The first two and the latter two components are interconnected via RF cables.
The APM network serves as the core spatial emulation component, precisely controlling the complex gain of each internal channel to reproduce the amplitude and phase variations across antenna array elements induced by the spatial positions and angular characteristics of the sensing targets. This network effectively transforms the limited number of RTS outputs into the spatially diverse signals required for large-scale antenna array testing. The RTS emulates the fundamental target characteristics including range, velocity, and RCS by introducing appropriate delay, Doppler shift, and complex attenuation to the sensing echo signals.

A  advantage of this framework architecture lies in its interface efficiency. The antenna ports of the ISAC BS, which may number in the hundreds for large-scale antenna arrays, connect exclusively to the APM network rather than directly interfacing with the RTS. This design significantly reduces the interface requirements on the RTS, whose RF ports are typically limited in number and costly to implement. The proposed framework operates in a conductive testing configuration, which represents a practical approach for initial commercial ISAC BS implementations. This methodology is particularly applicable to sub-6 GHz ISAC BSs, such as those operating at 4.9 GHz in China, where the first commercial deployments are expected and antenna connectors remain readily accessible for conducted testing procedures.
However, as ISAC technology evolves toward fully integrated antenna designs and migrates to higher millimeter-wave frequency bands, OTA testing methodologies will inevitably become necessary for future implementations, necessitating alternative approaches to the conducted testing framework presented herein.

It should be noted that the proposed framework in Fig. \ref{fig_framework} requires specific configuration according to the two sensing operation modes of the ISAC BS to determine the appropriate internal connections of the APM network. In the first mode, the antenna array operates for both transmission and reception in a duplex manner, referred to as the ADTR mode. In the second mode, the array is spatially partitioned into two distinct sub-arrays, with one dedicated to transmission and the other to reception \cite{umar2025possibilities, liu2025fundamental}, designated as the SATR mode \cite{smida2024band}.
The two sensing work modes correspond to different ISAC waveforms and serve to sense targets at varying distances, as detailed below:

1) The continuous wave (CW) based ISAC signal corresponds to the SATR work mode, used for sensing nearby targets \cite{zhang2021modulation,liu2021asynchronous}. In this working mode, the ISAC signal is transmitted by the transmit (Tx) antenna array, while the receive  (Rx) antenna array is ready to receive the echo signal in real-time. Since transmission and reception occur simultaneously, this mode eliminates near-field blind spots. However, the transmission power must be limited to avoid excessive self-interference between Tx and Rx \cite{xiao2022waveform, ma2021frac}, which restricts the ability to sense distant targets. The frequency-division multiplexing-based ISAC signal is a typical example of this sensing mode \cite{zhang2024target}.

2) The pulse wave (PW) based ISAC signal corresponds to the antenna arrays for ADTR work mode \cite{cai2023pulse}. The transmitter emits signals in short, high-power pulses, and the receiver activates in the transmission gap. Transmission and reception are time-division multiplexed, ensuring no interference between them \cite{liao2025pulse}. However, this approach results in near-field blind spots. Due to the higher transmission power, distant targets can be detected. The linear frequency modulation pulse signal is an example of this sensing mode.

Note that a single ISAC BS can operate in both aforementioned sensing modes and may dynamically switch between different operating modes across separate time slots within the same frame. This operational flexibility represents a significant design consideration under active development for commercial ISAC BS implementations.
Moreover, the sensing targets associated with the two modes may exhibit overlapping ranges.
Based on the two ISAC BS work modes described above, the proposed multi-target sensing emulation framework has two distinct configurations, which will be detailed in the following sub-sections.

\begin{figure}[!t]
	\centering
	\subfloat[]{\includegraphics[width=0.48\textwidth]{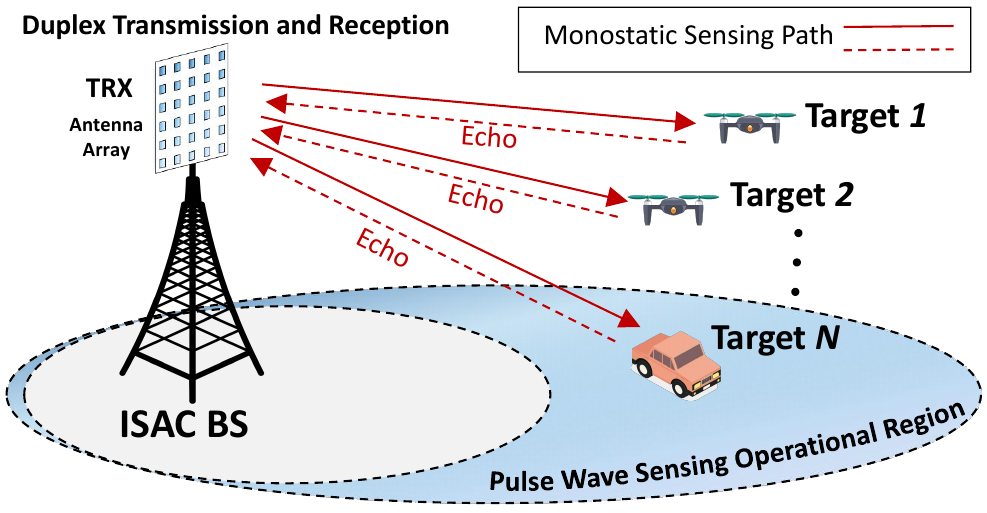}}
	\hfill 	
	\subfloat[]{\includegraphics[width=0.49 \textwidth]{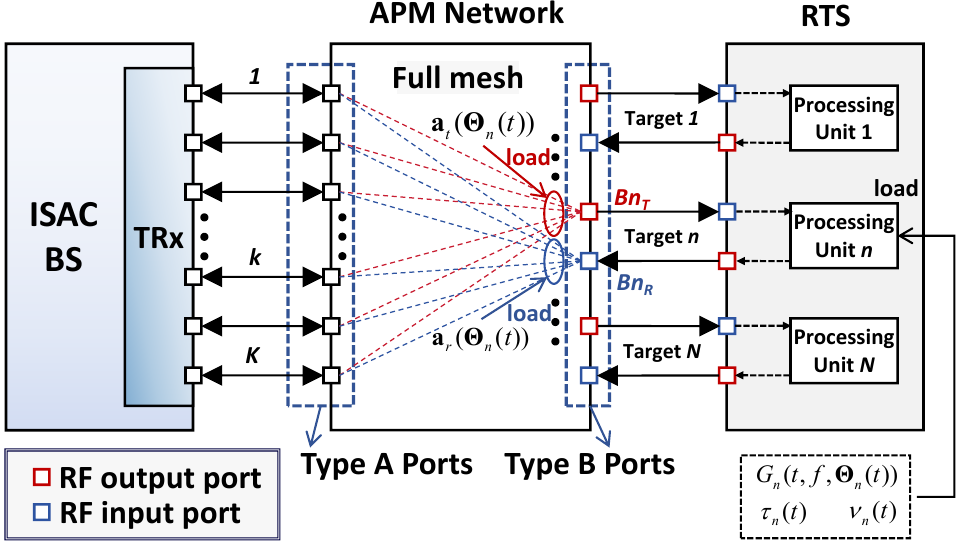}}
	\caption{ISAC BS multi-target sensing scenario in ADTR mode and the proposed emulation configuration  for this scenario. (a) Illustration of the sensing scenario. (b) Diagram of the emulation configuration. Note that in (b), only the schematic related to the 
		$n$-th target  emulation is shown for clarity.	}
	\label{fig_DTR}
\end{figure}

\subsection{Array Duplex Transmission and Reception Mode}
\subsubsection{Signal Model}
Consider an ISAC scenario as shown in Fig.~\ref{fig_DTR} (a), where an ISAC BS equipped with a uniform planar array (UPA) of $K$ antennas operates in ADTR mode. 
The ISAC BS senses $N$ distant targets using pulsed sensing waveforms via line-of-sight (LOS) propagation channels under a monostatic sensing setup.
Therefore, the channel frequency response (CFR) $\mathbf{H}^{D} (t,f)\in \mathbb{C}^{K \times K}$ consists of $N$ LoS channel components, which can be written as:
\begin{equation} 
	\mathbf{H}^{D} (t,f)=\sum_{n=1}^{N} \mathbf{H}^{D}_{n} (t,f).
\end{equation}
Since the ISAC BS typically senses distant targets in pulse waveform sensing mode, the far-field assumption is adopted. Thus, the $n$-th LoS channel component  can be written as:

\begin{align} \label{eq_2}
	 \mathbf{H}_{n}^{D} (t,f) =  & \underbrace{G_{n}(t, \mathbf{\Theta}_{n}(t)) \cdot \operatorname{exp}(\mathrm{j} 2\pi \nu_{n}(t) t )\cdot \operatorname{exp}(\mathrm{j} 2\pi f \tau_{n}(t))}_{\text{Emulated by the RTS in the Fig. 2(b)}}   \nonumber
	 \\
	&\cdot \underbrace{\mathbf{a}_{r}(\mathbf{\Theta}_{n}(t)) \cdot \mathbf{a}_{t}^{T}(\mathbf{\Theta}_{n}(t))}_{\text{Emulated by the  APM network in the Fig. 2(b)}}, 
\end{align}
where $\mathbf{\Theta}_{n}(t)$ is the spatial direction vector of $n$-th sensing point target at time $t$.  $G_{n}(t, \mathbf{\Theta}_{n}(t))$ is the channel gain of the $n$-th channel component, determined by path loss, antenna pattern, and RCS. Note that the inconsistency in the antenna patterns between the $N$ antennas is neglected. 
$\mathbf{a}_{r}(\mathbf{\Theta}_{n}(t)) \in \mathbb{C}^{K \times 1}$ and $\mathbf{a}_{t}(\mathbf{\Theta}_{n}(t)) \in \mathbb{C}^{K \times 1}$ are the Rx and Tx array steering vectors in the direction of $\mathbf{\Theta}_{n}(t)$, respectively.
Consider the ISAC BS operating in ADTR mode, where $\mathbf{a}_{r}(\mathbf{\Theta}_{n}(t))$ and $\mathbf{a}_{t}(\mathbf{\Theta}_{n}(t))$ are identical. Readers can refer to \cite{haupt2010antenna}  for the formulation of the steering vector of the UPA array.
The Doppler shift $\nu_{n}(t)$ and delay $\tau_{n}(t)$ of the $n$-th channel component at time $t$ are determined by the target velocity and range, respectively.

\subsubsection{Configuration}
For the ISAC BS operating in ADTR mode, the detailed configuration of the proposed framework is illustrated in Fig.~\ref{fig_DTR}~(b).
The RF ports of the APM network are divided into two groups interconnected by internal links with adjustable amplitude and phase. One group is designated as Type-A ports, while the other is denoted as Type B ports. 
For the scenario in Fig.~\ref{fig_DTR} (a), $K$ Type-A ports and $2N$ Type-B ports are enabled, forming a full-mesh connection such that the $k$-th Type-A port is connected to all $2N$ Type-B ports, and vice versa.
The $K$ antenna ports of the ISAC BS are connected to the $K$ Type-A ports of the APM network through RF cables in a one-to-one manner.
In this configuration, all Type-A ports support bidirectional signal flow, while Type-B ports are unidirectional. 
There are $N$ groups in the Type-B ports, each consisting of one Tx port and one Rx port. This configuration is determined by the structure of the RTS. Specifically, a RF input port on an RTS processing unit first acquires the signal, which is then processed internally, and subsequently retransmitted through the corresponding output port on the same unit.
In the $n$-th port group, the output Type-B port of the APM network delivers  the signal to the  RF input port of the $n$-th RTS processing unit via an RF cable. We use $B n_{T}$ as a common index to represent these two ports. 
 The signal then flows from the RF output port of the $n$-th RTS processing unit to the input Type-B port of the amplitude and phase modulation network, denoted by the common index $Bn_{R}$.

For the emulation of the $n$-th sensing target, the Tx array steering vector $\mathbf{a}_{t}(\mathbf{\Theta}_{n} (t))$ in (\ref{eq_2}) needs to be loaded into the APM channels linking the $K$ Type-A ports and the $B n_{T}$-th Type-B port. Similarly, the Rx array steering vector $\mathbf{a}_{r}(\mathbf{\Theta}_{n}(t))$ in (\ref{eq_2}) needs to be loaded into the APM channels linking the $K$ Type-A ports and the $B n_{R}$-th Type-B port.
Meanwhile, the channel gain $G_{n}(t, \mathbf{\Theta}_{n}(t))$, the Doppler shift $\nu_{n}(t)$, and the delay $\tau_{n}(t)$ are loaded into the $n$-th processing unit of the  RTS.
Finally, the spatio-temporal–frequency characteristics of the $n$-th target are emulated. The spatial characteristics are modeled by the APM, while the velocity and range characteristics are modeled by the RTS. Together, they jointly reproduce the features of $N$ distinct targets.

\begin{figure}[!t]
	\centering
	\subfloat[]{\includegraphics[width=0.48\textwidth]{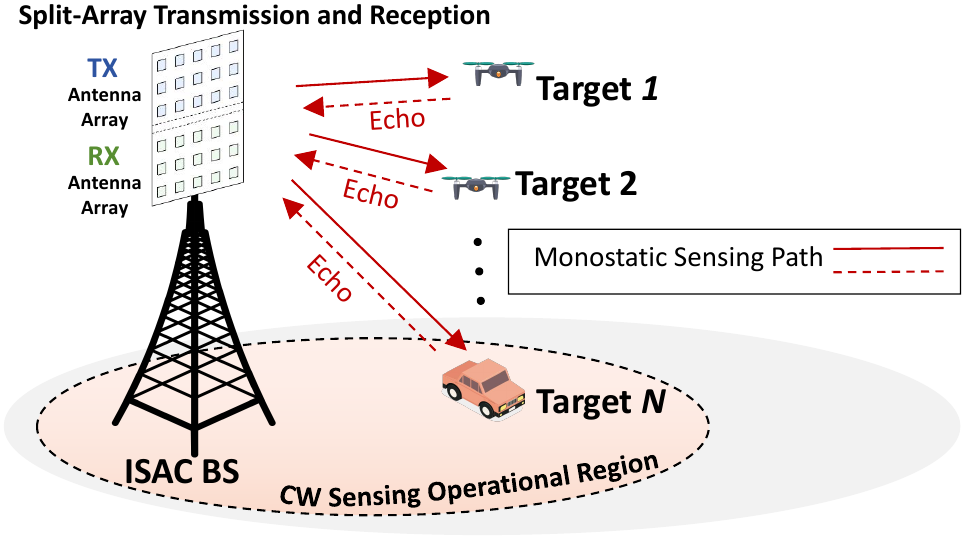}}
	\hfill 	
	\subfloat[]{\includegraphics[width=0.48 \textwidth]{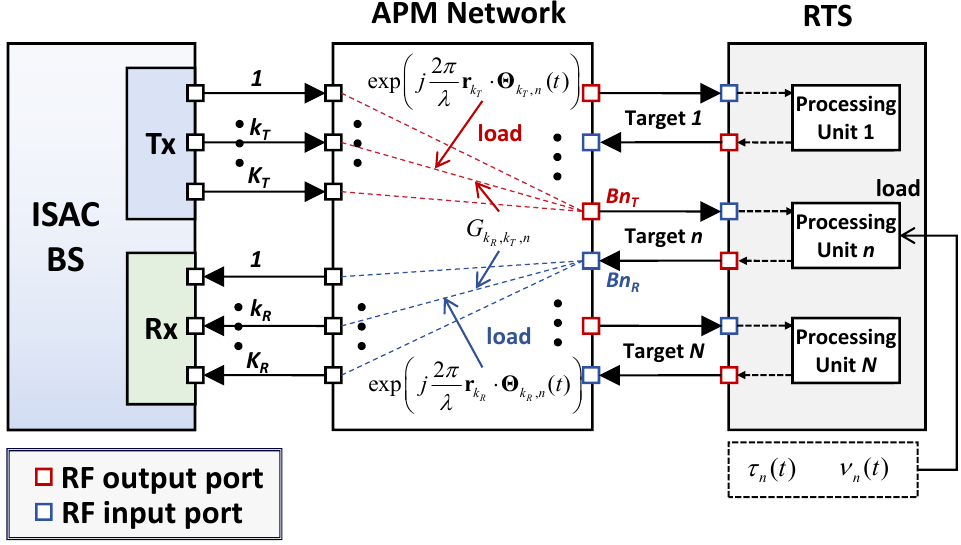}}
	\caption{ISAC BS multi-target sensing scenario in SATR mode and the proposed emulation configuration  for this scenario. (a) Illustration of the sensing scenario. (b) Diagram of the emulation configuration. Note that in (b), only the schematic related to the 
		$n$-th target  emulation is shown for clarity.}
	\label{fig_SATR}
\end{figure}

\subsection{Split-Array Transmission and Reception  Mode}
\subsubsection{Signal Model} 
Consider a sensing scenario of an ISAC BS operating in SATR mode, as shown in Fig.~\ref{fig_SATR} (a).
The  ISAC UPA BS employs the upper subarray with $K_T$ antennas to transmit CW-based ISAC signals, while the lower subarray with $K_R$ antennas receives sensing signals for detecting $N$ nearby targets.
Therefore, the CFR $\mathbf{H}^{S}(t,f) \in \mathbb{C}^{K_R \times K_T}$ between the $K_T$ Tx antennas and the $K_R$ Rx antennas for this SATR mode can be written as
\begin{equation} 
	\mathbf{H}^{S} (t,f)=\sum_{n=1}^{N} \mathbf{H}^{S}_{n} (t,f).
\end{equation}
Here, $\mathbf{H}^{S}_{n} (t,f) =\left \{ {h^{S}_{k_R,k_T,n}}(t,f)  \right \} \in \mathbb{C}^{K_R \times K_T}$  is  the $n$-th channel component  corresponding to the $n$-th sensing target. 
In this sensing mode, the target may be nearby, causing the far-field assumption to no longer hold. Therefore, the channel coefficient from the $k_T$-th Tx antenna to the $k_R$-th Rx antenna for $n$-th sensing target can be written as
\begin{align} \label{eq_4}
{h^{S}_{k_R,k_T,n}}&(t,f)=\underbrace{G_{k_R,k_T,n}(t,\mathbf{\Theta}_{k_R,n}(t), \mathbf{\Theta}_{k_T,n}(t))}_{\text{Emulated by the APM network in the Fig. 3(b)}}  \nonumber
\\
&\cdot \overbrace{\operatorname{exp}(\mathrm{j} \frac{2\pi}{\lambda} \mathbf{r}_{k_R}  \cdot  \mathbf{\Theta}_{k_R,n}(t))
\cdot\operatorname{exp}(\mathrm{j} \frac{2\pi}{\lambda} \mathbf{r}_{k_T}  \cdot  \mathbf{\Theta}_{k_T,n}(t))} \nonumber
\\
&\cdot \underbrace{ \operatorname{exp}(\mathrm{j} 2\pi \nu_{n}(t) t )\cdot \operatorname{exp}(\mathrm{j} 2\pi f \tau_{n}(t))}_{\text{Emulated by the RTS in the Fig. 3(b)}},
\end{align}
where $\mathbf{\Theta}_{k_R,n}(t)$ and $\mathbf{\Theta}_{k_T,n}(t)$ are the spatial direction vectors of the $n$-th sensing target relative to the $k_R$-th Rx antenna and the $k_T$-th Tx antenna at time $t$, respectively.
$G_{k_R,k_T,n}(t, \mathbf{\Theta}_{k_R,n}(t), \mathbf{\Theta}_{k_T,n}(t))$ is the channel gain determined by path loss, target spatial location, antenna pattern, and RCS.
Relative to the array phase center, $\mathbf{r}_{k_R}$ and $\mathbf{r}_{k_T}$ are the spatial location vectors of the $k_R$-th Rx antenna and the $k_T$-th Tx antenna, respectively.

\subsubsection{Configuration} 
For the ISAC BS operating in SATR  mode, the configuration of the proposed framework is illustrated in Fig.~\ref{fig_SATR} (b).
The physical cable connection method is similar to that in Fig.~\ref{fig_DTR} (b). However, due to the signal flow direction caused by the array transmission and reception mode, the APM network configuration is different.
The $K_T$ Type-A APM ports, which connect to the Tx antenna ports of the ISAC BS via cables, are only connected to the $B n_{T}$-th Type-B port ($n = 1, \dots, N$) through adjustable amplitude and phase internal  links, in order to load $\operatorname{exp} \left( j \frac{2\pi}{\lambda} \mathbf{r}_{k_T} \cdot \mathbf{\Theta}_{k_T,n}(t) \right)$ in (\ref{eq_4}).
Similarly, the $K_R$ Type-A APM ports are only connected to the $B n_{R}$-th Type-B port ($n = 1, \dots, N$) through adjustable amplitude and phase internal links in the APM network, thereby  loading $\operatorname{exp} \left( j \frac{2\pi}{\lambda} \mathbf{r}_{k_R} \cdot \mathbf{\Theta}_{k_R,n}(t) \right)$ in (\ref{eq_4}).
The Doppler shift $\nu_{n}(t)$ and the delay $\tau_{n}(t)$ of the $n$-th channel component are loaded into the $n$-th processing unit of the RTS.
For the $n$-th channel gain $G_{k_R,k_T,n}$, the target RCS and antenna element patterns may result in distinct responses across different antenna ports. This port-specific component is emulated by the APM network in conjunction with the spatial phase terms $\operatorname{exp} \left( j \frac{2\pi}{\lambda} \mathbf{r}_{k_R} \cdot \mathbf{\Theta}_{k_R,n}(t) \right)$ and $\operatorname{exp} \left( j \frac{2\pi}{\lambda} \mathbf{r}_{k_T} \cdot \mathbf{\Theta}_{k_T,n}(t) \right)$. When near-field effects are neglected, these spatial phase components can be directly emulated by the RTS.

\subsection{Discussion}
The proposed framework demonstrates scalability through the flexible amplitude and phase adjustment of the APM network across its internal channels. This inherent flexibility enables the emulation of spatially dynamic scenarios, plane wave propagation, and both near-field and far-field conditions through appropriate configuration of the amplitude-phase control parameters. The framework exhibits significant cost-effectiveness, as its fundamental design principle reduces RTS resource requirements. Specifically, the RTS resource consumption scales exclusively with the number of emulated targets and remains independent of the DUT antenna port count, representing a substantial advantage for large-scale antenna array testing.
A primary limitation of this framework stems from its conductive testing requirements. Establishing cable connections between the ISAC BS antenna ports and the APM network ports may present practical challenges for systems incorporating large-scale antenna arrays. While this approach remains well-suited for sub-6 GHz ISAC BSs where conductive access is readily available, OTA testing appears inevitable for future fully integrated system architectures. Furthermore, maintaining stringent inter-channel amplitude and phase consistency within the APM network is critical to ensure accurate sensing target emulation performance.

\begin{figure}[!t]
	\centering
	{\includegraphics[width=0.48\textwidth]{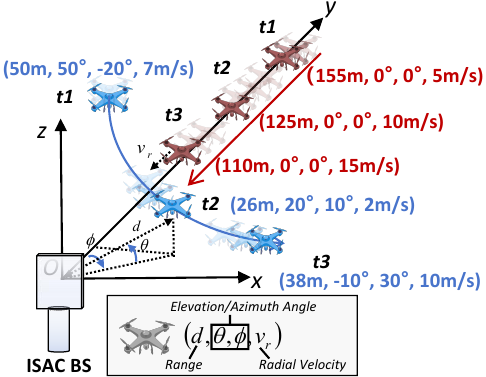}} 
	\caption{Target sensing scenario and coordinate system configuration.  We define the ISAC BS array center as the origin, with the vertical array plane facing the positive $y$-axis, and the $z$-axis pointing upward. The positive direction of the azimuth angle $\phi$ increases along the positive $x$-axis from the positive $y$-axis. The elevation angle $\theta$ increases in the direction of the positive $z$-axis from the $xoy$-plane.}
	\label{fig_targetADTR} 
\end{figure}

\begin{figure}[!t]
	\centering
	\subfloat[]{\includegraphics[width=0.48\textwidth]{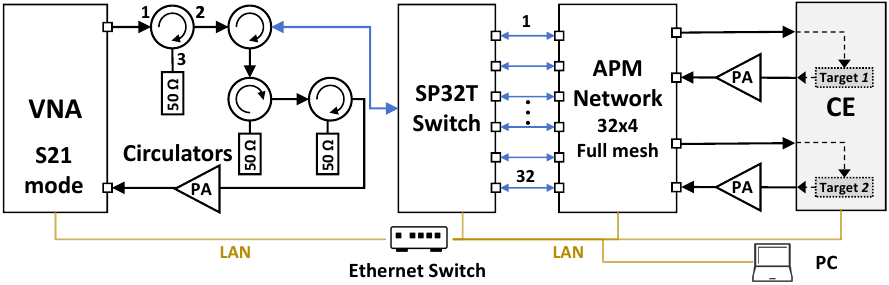}}
	\hfill 	
	\subfloat[]{\includegraphics[width=0.48 \textwidth]{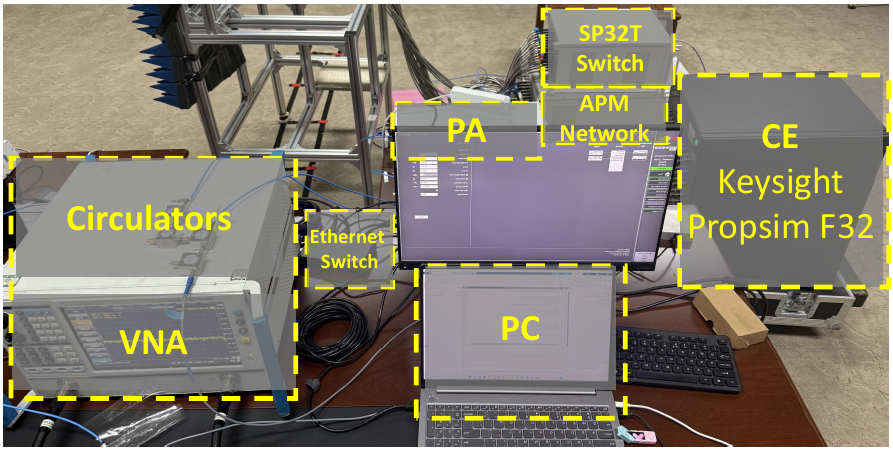}}
	\caption{Experiment setup of the sensing targets emulation for the ADTR mode of the ISAC BS. (a) Illustration of the measurement setup. (b) Photo of the setup in the laboratory condition.}
	\label{setupADTR}
\end{figure}

\section{Experimental Validation}
In this section, to validate the proposed framework under two distinct sensing operation modes, two representative drone target sensing scenarios are designed.

\subsection{ADTR mode of the ISAC BS}

\subsubsection{Target Sensing Scenario}
A dynamic scenario is designed in which an ISAC BS detects two low-altitude drone targets, as illustrated in Fig. \ref{fig_targetADTR}.
In three channel snapshots $t_1$, $t_2$, and $t_3$, the two drones have different range, elevation angle $\theta$, azimuth angle $\phi$, and radial velocity relative to the BS.
Drone 1 (blue) follows a curved trajectory, while Drone 2 (red) approaches the ISAC BS along the negative y-axis from a distant initial position.
The parameters of this target  sensing scenario are summarized in Table \ref{tab:drone_measurements}.

\begin{table*}[htbp]
	\centering
	\caption{Target and Measured Parameters of Drones at Different Channel Snapshots}
	\label{tab:drone_measurements}
	\begin{tabular}{@{}lc*{6}{c}@{}}
		\toprule
		\multirow{2}{*}{\textbf{Drone index}} & \multirow{2}{*}{\textbf{Parameter}} & \multicolumn{2}{c}{\textbf{Snapshot} $t_1$} & \multicolumn{2}{c}{\textbf{Snapshot} $t_2$} & \multicolumn{2}{c}{\textbf{Snapshot} $t_3$} \\
		\cmidrule(lr){3-4}\cmidrule(lr){5-6}\cmidrule(lr){7-8}
		& & \cellcolor{gray!1} \textbf{Target} & \cellcolor{gray!15} \textbf{Measured} &\cellcolor{gray!1} \textbf{Target} &  \cellcolor{gray!15} \textbf{Measured} &\cellcolor{gray!1} \textbf{Target} & \cellcolor{gray!15} \textbf{Measured} \\
		\midrule
		\multirow{5}{*}{Drone 1} & Range ($m$) & \cellcolor{gray!1}50 & \cellcolor{gray!15}50 & \cellcolor{gray!1}26 & \cellcolor{gray!15}26 & \cellcolor{gray!1}38 & \cellcolor{gray!15}38 \\
		& Radial Velocity ($m/s$) & \cellcolor{gray!1}7 & \cellcolor{gray!15}7 & \cellcolor{gray!1}2 & \cellcolor{gray!15}2 & \cellcolor{gray!1}10 & \cellcolor{gray!15}10 \\
		& Elevation Angle $\theta$ (°) & \cellcolor{gray!1}50 & \cellcolor{gray!15}50 & \cellcolor{gray!1}20 & \cellcolor{gray!15}20 & \cellcolor{gray!1}-10 & \cellcolor{gray!15}-10 \\
		& Azimuth Angle $\phi$ (°) & \cellcolor{gray!1}-20 & \cellcolor{gray!15}-20 & \cellcolor{gray!1}10 & \cellcolor{gray!15}10 & \cellcolor{gray!1}30 & \cellcolor{gray!15}30 \\
		& Normalized Channel Gain (dB) & \cellcolor{gray!1}-5 & \cellcolor{gray!15}-5.1 & \cellcolor{gray!1}0 & \cellcolor{gray!15}0 & \cellcolor{gray!1}-3 & \cellcolor{gray!15}-3 \\
		\midrule
		\multirow{5}{*}{Drone 2} & Range ($m$) & \cellcolor{gray!1}155 & \cellcolor{gray!15}155 & \cellcolor{gray!1}125 & \cellcolor{gray!15}125 & \cellcolor{gray!1}110 & \cellcolor{gray!15}110 \\
		& Radial Velocity ($m/s$) & \cellcolor{gray!1}5 & \cellcolor{gray!15}5 & \cellcolor{gray!1}10 & \cellcolor{gray!15}10 & \cellcolor{gray!1}15 & \cellcolor{gray!15}15 \\
		& Elevation Angle $\theta$ (°) & \cellcolor{gray!1}0 & \cellcolor{gray!15}0 & \cellcolor{gray!1}0 & \cellcolor{gray!15}0 & \cellcolor{gray!1}0 & \cellcolor{gray!15}0 \\
		& Azimuth Angle $\phi$ (°) & \cellcolor{gray!1}0 & \cellcolor{gray!15}0 & \cellcolor{gray!1}0 & \cellcolor{gray!15}0 & \cellcolor{gray!1}0 & \cellcolor{gray!15}0 \\
		& Normalized Channel Gain (dB) & \cellcolor{gray!1}-25 & \cellcolor{gray!15}-25.9 & \cellcolor{gray!1}-20 & \cellcolor{gray!15}-20.7 & \cellcolor{gray!1}-13 & \cellcolor{gray!15}-12.8 \\
		\bottomrule
	\end{tabular}
\end{table*}
\begin{figure*}[!t]
	\centering
	{\includegraphics[width=0.90\textwidth]{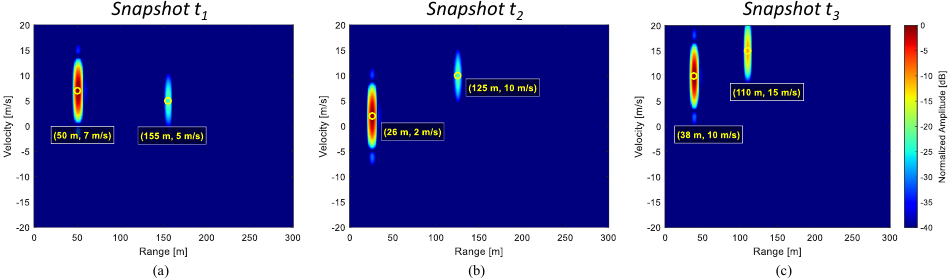}} 
	\caption{Estimated joint range and velocity profiles at different time snapshots. (a) Snapshot $t_1$. (b) Snapshot $t_2$. (c) Snapshot $t_3$.}
	\label{fig_result_range_doppler} 
\end{figure*}

\begin{table}
	\caption{Measurement Equipment and Parameters}
	\label{Mea parameter1}
	\centering
	\renewcommand\arraystretch{1.1}
	\begin{tabular}{@{}ll@{}}
		\toprule
		\textbf{Equipment/Parameter}          & \textbf{Value}         
		\\ 
		\midrule
		\underline{CE}             & Keysight Propsim F32    \\
		CE bandwidth                 & 40 MHz                  \\
		CE center frequency                    & 3500 MHz                  \\ 
		\underline{VNA}                          & Ceyear 3671C     \\
		VNA center frequency                  & 3500 MHz                \\
		VNA bandwidth                         & 40 MHz                  \\
		VNA IF bandwidth                         & 2000 Hz                  \\
		Number of frequency points  $N_f$           & 1001                   \\
		VNA Power                         & 20 dBm               \\
		\underline{Power Amplifier}              & General Test PA M80706A 
		\\ 
		\underline{Switch}              & Topyoung TSS-C601 
		\\ 
		\underline{APM network}              & Topyoung MCS2350A2-64B16
		\\
		\bottomrule
	\end{tabular}
\end{table}

\subsubsection{Experiment Setup}
To validate the proposed framework for ADTR mode, the experimental configuration is depicted in Fig. \ref{setupADTR}, which includes both the  ISAC BS and the testing system:

$\bullet$ To emulate an ISAC BS operating in the ADTR sensing mode, a vector network analyzer (VNA), a single-pole 32-throw (SP32) switch, and four cascaded circulators are employed to ensure high transceiver isolation. This configuration emulates an ISAC BS equipped with a 4$\times$8 UPA operating at 3.5 GHz with half-wavelength element spacing. The high transceiver isolation provided by the cascaded circulator configuration enables simultaneous transmission and reception signal recording, thereby accurately mimicking the ISAC BS ADTR mode operation. This isolation capability is critical for ADTR mode emulation, as it allows the system to maintain signal integrity while operating in full-duplex sensing mode. The VNA and switching system are utilized to acquire the sensing channel matrix, with subsequent array signal processing performed through offline post-processing algorithms.

$\bullet$ 
For the emulation system, the APM network was configured with a 4$\times$32 full-mesh architecture. The Keysight PROPSIM F32 CE was employed as the RTS, which has been validated in \cite{wang2025channel}. The system was configured to emulate two targets utilizing four RF ports while consuming only two internal channel resources.
As mentioned in Section II, the Rx array steering vector $\mathbf{a}_{r}(\mathbf{\Theta}_{n}(t))$ in (\ref{eq_2}) is loaded into the APM network.
And the channel gain $G_{n}(t, \mathbf{\Theta}_{n}(t))$, the Doppler shift $\nu_{n}(t)$, and the delay $\tau_{n}(t)$ are loaded into the CE.
 The channel impulse response (CIR) files loaded into the CE are configured for Drone 1 and Drone 2 with corresponding delays, power levels, and radial velocities. 
 The update rate for these CIR  files is set to twice the maximum Doppler frequency, with a time duration of  $N_t = 1000$ CIR  samples for each snapshot.
For velocity estimation, the CE pauses at each CIR sample. After the VNA measures the corresponding CFR over $N_f = 1001$ frequency points with $40$~MHz bandwidth, the CE advances to the next CIR sample and repeats the process until the sequence is complete\footnote{It is well established that VNA can only measure quasi-static CFRs. However, in this measurement approach, time-variant channels are preloaded into the CE, allowing the system to step through and pause at specific channel snapshots for VNA measurements. This operational methodology is widely adopted for multiple-input multiple-output OTA channel validation measurements \cite{3gpp-tr-38-827}.}.
 Consequently, a  CFR dataset of dimensions $N_t \times N_f \times 32$ is recorded.
All instruments are connected via Ethernet and controlled through a personal computer (PC).
Finally, the detailed  instrument parameters for this experiment are summarized in Table \ref{Mea parameter1}.

\begin{figure*}[!t]
	\centering
	{\includegraphics[width=0.95\textwidth]{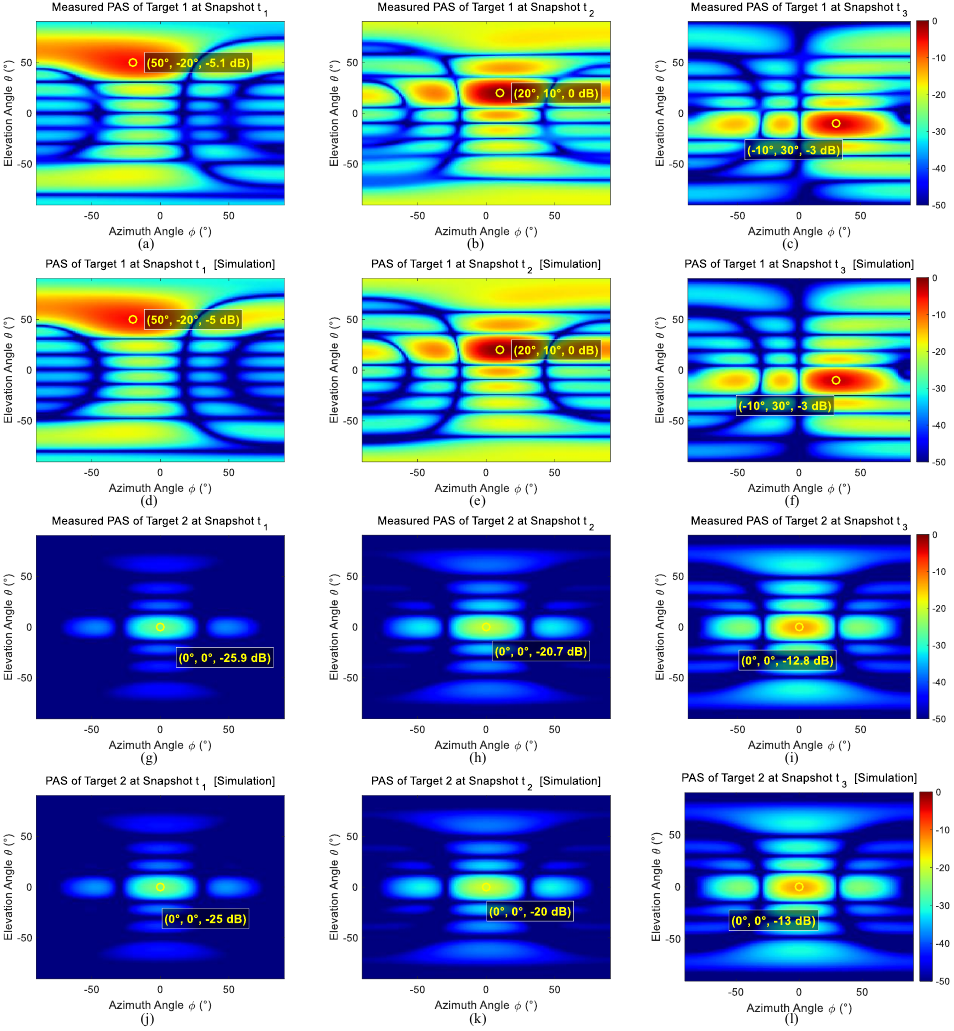}} 
	\caption{The measured and simulated PAS of target 1 and target 2 at different snapshots. (a)-(c) Measured PAS of target 1 at three snapshots. (d)-(f) Simulated PAS of target 1 at three snapshots. (g)-(i) Measured PAS of target 2 at three snapshots. (j)-(l) Simulated PAS of target 2 at three snapshots.}
	
	\label{fig_PAS} 
\end{figure*}

\subsubsection{Result Analysis}
For the estimation of the three-dimensional parameters of the sensing targets, including angle, range, and velocity, a two-dimensional Fourier transform (FT) is first applied to the joint time–frequency data of dimension $N_t \times N_f$ obtained from a randomly selected antenna at each snapshot.
This yields the joint Doppler–delay domain data, which is subsequently mapped to the joint range–velocity domain as illustrated in Fig. \ref{fig_result_range_doppler}.
For joint range–velocity profile of each snapshot, two peak values are marked with yellow circles, from which the estimated velocities and ranges of each sensing target are obtained.
It can be observed that the velocities and ranges of drone 1 and drone 2 are well emulated and estimated for each snapshot, consistent with the values set in the target scenario.
The stronger echo power from drone 1 results in a darker peak intensity in the figure.

Next, we performed a FT on the frequency data of dimension $N_f \times 1$ at each snapshot for each of the 32 antenna ports to obtain the delay domain data. We then applied a 2D beamforming algorithm to the delay domain data from the 32 antenna ports to derive the power-angular-delay profile (PADP) at each snapshot. Finally, based on the estimated delay of each drone, we sliced the PADP at each snapshot to obtain the measured power-angular-spectrum (PAS) for each drone, which was compared with the theoretical simulated PAS, as shown in Fig. \ref{fig_PAS}.
The peak of each sub-figure is marked with a yellow circle, and the corresponding estimated values of $\theta$, $\phi$, and power are provided.
From the estimated values, the measured results based on the proposed framework are highly similar to the values set in the target scenario. The measured estimated $\theta$ and $\phi$ are highly consistent, while the power shows a slight discrepancy, with the maximum difference occurring in snapshot $t_1$ of target 2, where there is an approximately 0.9 dB difference, as shown in Fig. \ref{fig_PAS} (g). 
In addition to the peaks, the overall PAS pattern, including the main lobe and sidelobe profiles, is also essential for evaluation, as it determines the sensing performance. From the figure, it can be observed that the measured PAS is highly similar to the theoretical simulated PAS within the 50 dB dynamic range for each target at each snapshot, both in terms of the main lobe profile and the sidelobes. In certain low-power regions of the PAS, the measured PAS shows some discrepancies with the target PAS, as shown in Fig. \ref{fig_PAS} (a)-(f).
This is primarily due to the limited number of phase adjustment bits in the APM.
Finally, all the experimentally measured parameters are summarized along with the target parameters in Table \ref{tab:drone_measurements}, where the cells corresponding to the measured values are highlighted with a darker background to enhance readability.

\begin{figure}[!t]
	\centering
	{\includegraphics[width=0.48\textwidth]{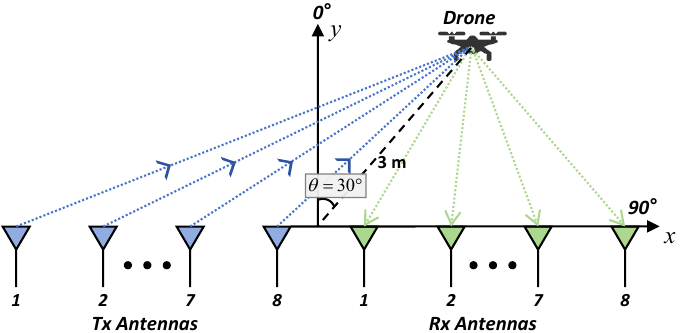}} 
	\caption{Target sensing scenario for the  ISAC ULA BS  operating in  the SATR mode, along with the coordinate system configuration.
		The $y$-axis is perpendicular to the array plane and points in the nominal transmission direction. The $x$-axis is parallel to the array aperture, with the positive $x$-axis pointing from the Tx sub-array to the Rx sub-array. The angle $\theta$ is measured from the positive $y$-axis toward the positive $x$-axis in the counterclockwise sense, with $0^\circ$ aligned with the positive $y$-axis and $90^\circ$ aligned with the positive x-axis.
	}
	\label{fig_SATR_Scen} 
\end{figure}

\subsection{SATR Mode of the ISAC BS} 
 \subsubsection{Target Sensing Scenario}
We consider a two-dimensional static sensing scenario where an ISAC BS equipped with a 1$\times$16 uniform linear array (ULA) senses a drone in the SATR mode, as shown in Fig.~\ref{fig_SATR_Scen}. The ULA operates at 3.5 GHz with a half-wavelength spacing, using the first eight antennas for signal transmission and the remaining eight antennas for signal reception.
The sensing target, a drone, is modeled as a point target located at $\theta = 30^\circ$ and positioned 3 meters from the center of the ULA.

\begin{figure}[!t]
	\centering
	\subfloat[]{\includegraphics[width=0.48\textwidth]{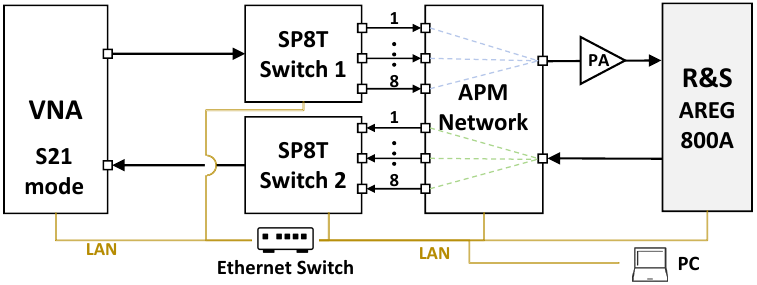}}
	\hfill 	
	\subfloat[]{\includegraphics[width=0.48 \textwidth]{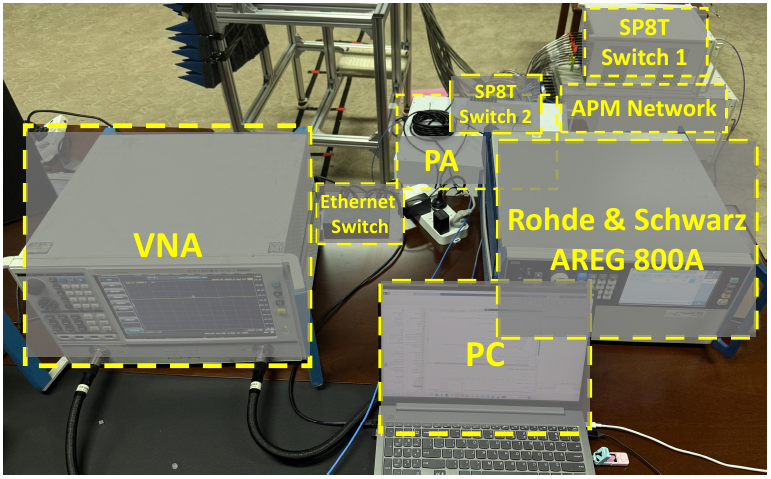}}
	\caption{Experiment setup of the sensing targets emulation for the SATR mode of the ISAC BS. (a) Illustration of the measurement setup. (b) Photo of the setup in the laboratory condition.}
	\label{setup_SATR}
	\vspace{-10pt}
\end{figure}

\subsubsection{Experiment Setup}
To validate the capability of the proposed framework for an ISAC BS operating in the SATR mode, the designed experimental setup is illustrated in Fig.~\ref{setup_SATR} (a) and (b).
In this setup, two single-pole eight-throw (SP8T) switches and a VNA are used to emulate a  ISAC BS equipped with a 1$ \times $16 ULA in the SATR mode. One switch emulates the eight Tx antennas, while the other switch emulates the eight Rx antennas. 
The APM network is enabled with 16 Type-A ports and 2 Type-B ports. The first eight Type-A ports are connected to Switch 1, while the remaining ones are connected to Switch 2. The APM network is configured with a total of 16 internal channels featuring adjustable amplitude and phase between its RF ports, as detailed in the connection diagram shown in Fig.~\ref{setup_SATR} (a).
The phase terms $\operatorname{exp} \left( j \frac{2\pi}{\lambda} \mathbf{r}_{k_R} \cdot \mathbf{\Theta}_{k_R,n}(t) \right)$ and $\operatorname{exp} \left( j \frac{2\pi}{\lambda} \mathbf{r}_{k_T} \cdot \mathbf{\Theta}_{k_T,n}(t) \right)$ in (\ref{eq_4}), which are calculated according to Fig.~\ref{fig_SATR_Scen}, are loaded into the APM network. 
We assume that the amplitude of the received sensing echo at the ULA antennas is uniform.
Additionally, the VNA setting remains consistent with the previous setup. We use the Rohde \& Schwarz AREG800A \cite{rohde_schwarz_2024_areg800a} as the RTS and import the range data of the target drone, which operates at 3.5 GHz with a 1 GHz bandwidth. All instruments are connected via Ethernet and controlled through a personal computer.

During the measurement, SP8T switch 1 is set to the first throw position while SP8T switch 2 sequentially cycles through all its throws. The VNA measures the CFR at each step. This process repeats until switch 1 has iterated through all its throws, resulting in a recorded CFR dataset with dimensions of $8 \times 8 \times N_f$.
\begin{figure}[!t]
	\centering
	{\includegraphics[width=0.48\textwidth]{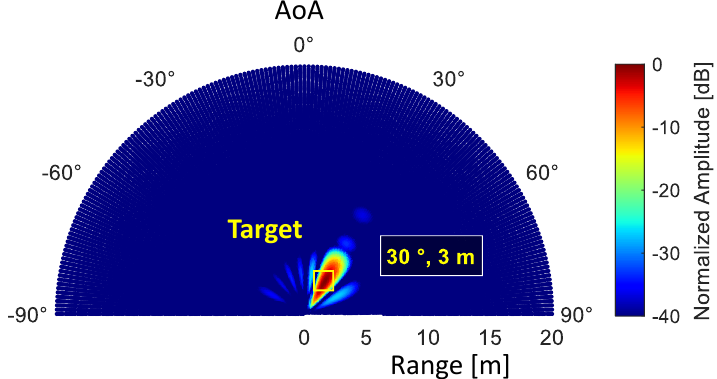}} 
	\caption{The polar plot of the estimated joint  angular and range profile. The peak and corresponding values are highlighted with yellow boxes and text.}
	\label{fig_SATR_re} 
		\vspace{-10pt}
\end{figure}
\subsubsection{Result Analysis}
Based on the recorded CFR dataset, a joint estimation in the range and angular domains was performed, as illustrated in Fig.~\ref{fig_SATR_re}.
The estimated AoA and range are fully consistent with the target set values, demonstrating that the proposed framework successfully emulates the sensing target for the SATR mode of the ISAC BS.
Note that for Fig.~\ref{fig_SATR_re}, the Hanning window is applied to suppress sidelobes.

\subsection{Discussion}
In this section, two experiments are conducted to validate the feasibility of the proposed framework for sensing target emulation across different ISAC BS operational modes. The emulated ISAC BS successfully estimated the sensing target characteristics, demonstrating close alignment with the predefined target scenario configurations. The designed experiments serve exclusively to validate the fundamental functionality of the proposed framework and do not comprehensively replicate the operational conditions of practical ISAC BS deployments.
The experimental design incorporates several simplifications that warrant acknowledgment. The algorithms employed for target angle, range, and velocity estimation in Fig. \ref{fig_targetADTR} utilize fundamental FT-based estimation approaches selected for measurement robustness and validation reliability. In practical ISAC BS implementations, more sophisticated parameter estimation algorithms, such as high-resolution sensing parameter estimators, would typically be employed to achieve enhanced estimation accuracy and resolution capabilities.
Furthermore, the measurements were conducted in step mode, wherein each data snapshot is individually captured and recorded in a sequential manner. In contrast, actual deployment scenarios require ISAC BSs and sensing targets to operate in continuous real-time interaction. Nevertheless, these experimental simplifications do not compromise the validation of the proposed framework's fundamental functionality and operational principles.
As previously discussed, ISAC BSs may dynamically transition between different sensing modes across distinct time slots within the same operational frame structure. The precise synchronization requirements among the APM network, the ISAC BS, and the RTS represent a practical challenge that must be addressed for successful real-world deployment.

	\vspace{-5pt}

\section{Conclusion}
This paper proposes a framework for multi-target sensing emulation in conducted testing of ISAC BS, incorporating an APM network and a RTS. Based on this framework, configurations are examined for different sensing operation modes of the ISAC BS. For experimental validation, two representative sensing scenarios and corresponding experiment setups are designed for the ADTR and the SATR modes.
In the first experiment, different AoAs, ranges, velocities, and power levels of two drones are accurately estimated by a emulated ISAC BS operating in ADTR mode, with a maximum power error of only 0.9 dB across three channel snapshots.  
In the second experiment, the range and velocity of a target drone are successfully emulated and estimated by a emulated ISAC BS operating in SATR mode.  
The feasibility of the proposed framework is demonstrated through these two experiments.  
Given the conducted connection limitations of the proposed framework, future work may explore OTA testing methods for ISAC systems.

\bibliographystyle{IEEEtran}
\addcontentsline{toc}{section}{\refname}
\bibliography{ref_ConductedISAC}

\begin{thebibliography}{10}
\providecommand{\url}[1]{#1}
\csname url@samestyle\endcsname
\providecommand{\newblock}{\relax}
\providecommand{\bibinfo}[2]{#2}
\providecommand{\BIBentrySTDinterwordspacing}{\spaceskip=0pt\relax}
\providecommand{\BIBentryALTinterwordstretchfactor}{4}
\providecommand{\BIBentryALTinterwordspacing}{\spaceskip=\fontdimen2\font plus
\BIBentryALTinterwordstretchfactor\fontdimen3\font minus
  \fontdimen4\font\relax}
\providecommand{\BIBforeignlanguage}[2]{{%
\expandafter\ifx\csname l@#1\endcsname\relax
\typeout{** WARNING: IEEEtran.bst: No hyphenation pattern has been}%
\typeout{** loaded for the language `#1'. Using the pattern for}%
\typeout{** the default language instead.}%
\else
\language=\csname l@#1\endcsname
\fi
#2}}
\providecommand{\BIBdecl}{\relax}
\BIBdecl

\bibitem{wang2023road}
C.-X. Wang, X.~You, X.~Gao, X.~Zhu, Z.~Li, C.~Zhang, H.~Wang, Y.~Huang,
  Y.~Chen, H.~Haas \emph{et~al.}, ``{On the road to 6G: Visions, requirements,
  key technologies, and testbeds},'' \emph{IEEE Communications Surveys \&
  Tutorials}, vol.~25, no.~2, pp. 905--974, 2023.

\bibitem{dong2022sensing}
F.~Dong, F.~Liu, Y.~Cui, W.~Wang, K.~Han, and Z.~Wang, ``Sensing as a service
  in {6G} perceptive networks: A unified framework for {ISAC} resource
  allocation,'' \emph{IEEE Transactions on Wireless Communications}, vol.~22,
  no.~5, pp. 3522--3536, 2022.

\bibitem{liu2022integrated}
F.~Liu, Y.~Cui, C.~Masouros, J.~Xu, T.~X. Han, Y.~C. Eldar, and S.~Buzzi,
  ``Integrated sensing and communications: Toward dual-functional wireless
  networks for {6G} and beyond,'' \emph{IEEE journal on selected areas in
  communications}, vol.~40, no.~6, pp. 1728--1767, 2022.

\bibitem{Chen2024}
G.~Chen, R.~Zhang, H.~Ren, X.~Lin, and W.~Wu, ``Joint beamforming design for
  dual-functional radar-communication systems under beampattern gain
  constraints,'' \emph{ZTE Communications}, vol.~22, no.~3, pp. 13--20, 2024.

\bibitem{wang2025channel}
Z.~Wang, C.~Li, H.~Sun, S.~Qiao, and W.~Fan, ``Channel emulator enabled
  over-the-air testing platform for integrated sensing and communication
  systems: Framework and experimental validation,'' \emph{IEEE Transactions on
  Instrumentation and Measurement}, 2025.

\bibitem{zhang2022time}
Q.~Zhang, H.~Sun, X.~Gao, X.~Wang, and Z.~Feng, ``Time-division isac enabled
  connected automated vehicles cooperation algorithm design and performance
  evaluation,'' \emph{IEEE Journal on Selected Areas in Communications},
  vol.~40, no.~7, pp. 2206--2218, 2022.

\bibitem{yang2024isac}
J.~Yang, H.~Que, T.~Du, L.~Liang, X.~Li, C.-K. Wen, and S.~Jin, ``Isac
  prototype system for multi-domain cooperative communication networks,''
  \emph{IEEE Wireless Communications Letters}, 2024.

\bibitem{koopman2016challenges}
P.~Koopman and M.~Wagner, ``Challenges in autonomous vehicle testing and
  validation,'' \emph{SAE International Journal of Transportation Safety},
  vol.~4, no.~1, pp. 15--24, 2016.

\bibitem{gadringer2018virtual}
M.~E. Gadringer, H.~Schreiber, A.~Gruber, M.~Vorderderfler, D.~Amschl,
  W.~B{\"o}sch, S.~Metzner, H.~Pfl{\"u}gl, and M.~Paulweber, ``Virtual reality
  for automotive radars,'' \emph{e \& i Elektrotechnik und
  Informationstechnik}, vol. 135, no.~4, pp. 335--343, 2018.

\bibitem{korner2021multirate}
G.~K{\"o}rner, M.~Hoffmann, S.~Neidhardt, M.~Beer, C.~Carlowitz, and
  M.~Vossiek, ``Multirate universal radar target simulator for an accurate
  moving target simulation,'' \emph{IEEE Transactions on Microwave Theory and
  Techniques}, vol.~69, no.~5, pp. 2730--2740, 2021.

\bibitem{zhai2024wideband}
W.~Zhai, R.~Wang, X.~Wang, F.~Shi, X.~Pang, Y.~Gao, and W.~Cui, ``Wideband
  photonic radar target simulator based on all-optical iq upconverter,''
  \emph{IEEE Transactions on Microwave Theory and Techniques}, 2024.

\bibitem{buddappagari2021over}
S.~Buddappagari, M.~E. Asghar, F.~Baumg{\"a}rtner, S.~Graf, F.~Kreutz,
  A.~L{\"o}ffler, J.~Nagel, T.~Reichmann, R.~Stephan, and M.~A. Hein,
  ``Over-the-air vehicle-in-the-loop test system for installed-performance
  evaluation of automotive radar systems in a virtual environment,'' in
  \emph{2020 17th European Radar Conference (EuRAD)}.\hskip 1em plus 0.5em
  minus 0.4em\relax IEEE, 2021, pp. 278--281.

\bibitem{asghar2021radar}
M.~E. Asghar, S.~Buddappagari, F.~Baumg{\"a}rtner, S.~Graf, F.~Kreutz,
  A.~L{\"o}ffler, J.~Nagel, T.~Reichmann, R.~Stephan, and M.~Hein, ``Radar
  target simulator and antenna positioner for real-time over-the-air
  stimulation of automotive radar systems,'' in \emph{2020 17th European Radar
  Conference (EuRAD)}.\hskip 1em plus 0.5em minus 0.4em\relax IEEE, 2021, pp.
  95--98.

\bibitem{scheiblhofer2017low}
W.~Scheiblhofer, R.~Feger, A.~Haderer, and A.~Stelzer, ``A low-cost
  multi-target simulator for fmcw radar system calibration and testing,'' in
  \emph{2017 European Radar Conference (EURAD)}.\hskip 1em plus 0.5em minus
  0.4em\relax IEEE, 2017, pp. 343--346.

\bibitem{gadringer2018radar}
M.~E. Gadringer, F.~M. Maier, H.~Schreiber, V.~P. Makkapati, A.~Gruber,
  M.~Vorderderfler, D.~Amschl, S.~Metzner, H.~Pfl{\"u}gl, W.~B{\"o}sch
  \emph{et~al.}, ``Radar target stimulation for automotive applications,''
  \emph{IET Radar, Sonar \& Navigation}, vol.~12, no.~10, pp. 1096--1103, 2018.

\bibitem{diewald2021arbitrary}
A.~Diewald, B.~Nuss, M.~Pauli, and T.~Zwick, ``Arbitrary angle of arrival in
  radar target simulation,'' \emph{IEEE Transactions on Microwave Theory and
  Techniques}, vol.~70, no.~1, pp. 513--520, 2021.

\bibitem{schoeder2021flexible}
P.~Schoeder, V.~Janoudi, B.~Meinecke, D.~Werbunat, and C.~Waldschmidt,
  ``Flexible direction-of-arrival simulation for automotive radar target
  simulators,'' \emph{IEEE Journal of Microwaves}, vol.~1, no.~4, pp. 930--940,
  2021.

\bibitem{schoeder2023unified}
P.~Schoeder, V.~Janoudi, T.~Grebner, and C.~Waldschmidt, ``A unified model of
  coherent direction-of-arrival simulation for radar target simulators,''
  \emph{IEEE Transactions on Aerospace and Electronic Systems}, vol.~59, no.~4,
  pp. 4738--4743, 2023.

\bibitem{zhang2020achieving}
F.~Zhang, W.~Fan, and Z.~Wang, ``Achieving wireless cable testing of high-order
  mimo devices with a novel closed-form calibration method,'' \emph{IEEE
  Transactions on Antennas and Propagation}, vol.~69, no.~1, pp. 478--487,
  2020.

\bibitem{rohde_schwarz_2024_areg800a}
\BIBentryALTinterwordspacing
{Rohde \& Schwarz}, \emph{{R\&S\textregistered{} AREG800A Automotive Radar Echo
  Generator — Specifications}}, Munich, Germany, March 2024, version 05.00,
  PD 3609.8015.22. [Online]. Available:
  \url{https://www.rohde-schwarz.com/cz/products/test-and-measurement/echo-generators/rs-areg800a-automotive-radar-echo-generator_63493-1044352.html}
\BIBentrySTDinterwordspacing

\bibitem{umar2025possibilities}
M.~Umar, M.~Ramzan, S.~Kamal, M.~S. Ahmad, and P.~Sen, ``Possibilities and
  challenges for a phased array antenna system in isac: A hardware
  perspective,'' \emph{IEEE Access}, 2025.

\bibitem{liu2025fundamental}
F.~Liu, T.~Zhang, Z.~Zhang, B.~Cao, Y.~Shen, and Q.~Zhang, ``Fundamental limits
  of pulse based uwb isac systems: A parameter estimation perspective,''
  \emph{IEEE Internet of Things Journal}, 2025.

\bibitem{smida2024band}
B.~Smida \emph{et~al.}, ``In-band full-duplex: The physical layer,''
  \emph{Proceedings of the IEEE}, vol. 112, no.~5, pp. 433--462, 2024.

\bibitem{zhang2021modulation}
H.~Zhang, T.~Zhang, and Y.~Shen, ``Modulation symbol cancellation for
  {OTFS}-based joint radar and communication,'' in \emph{2021 IEEE
  International Conference on Communications Workshops (ICC Workshops)}.\hskip
  1em plus 0.5em minus 0.4em\relax IEEE, 2021, pp. 1--6.

\bibitem{liu2021asynchronous}
F.~Liu, T.~Zhang, and P.~Cao, ``Asynchronous integration of communication and
  localization systems using {IR-UWB} signals,'' in \emph{MILCOM 2021-2021 IEEE
  Military Communications Conference (MILCOM)}.\hskip 1em plus 0.5em minus
  0.4em\relax IEEE, 2021, pp. 521--527.

\bibitem{xiao2022waveform}
Z.~Xiao and Y.~Zeng, ``Waveform design and performance analysis for full-duplex
  integrated sensing and communication,'' \emph{IEEE Journal on Selected Areas
  in Communications}, vol.~40, no.~6, pp. 1823--1837, 2022.

\bibitem{ma2021frac}
D.~Ma \emph{et~al.}, ``{FRaC}: Fmcw-based joint radar-communications system via
  index modulation,'' \emph{IEEE journal of selected topics in signal
  processing}, vol.~15, no.~6, pp. 1348--1364, 2021.

\bibitem{zhang2024target}
Z.~Zhang, H.~Ren, C.~Pan, S.~Hong, D.~Wang, J.~Wang, and X.~You, ``Target
  localization in cooperative isac systems: A scheme based on 5g nr ofdm
  signals,'' \emph{IEEE Transactions on Communications}, 2024.

\bibitem{cai2023pulse}
S.~Cai, L.~Chen, Y.~Chen, H.~Yin, and W.~Wang, ``Pulse-based isac: Data
  recovery and ranging estimation for multi-path fading channels,'' \emph{IEEE
  Transactions on Communications}, vol.~71, no.~8, pp. 4819--4838, 2023.

\bibitem{liao2025pulse}
Z.~Liao, F.~Liu, S.~Li, Y.~Xiong, W.~Yuan, C.~Masouros, and M.~Lops, ``Pulse
  shaping for random isac signals: The ambiguity function between symbols
  matters,'' \emph{IEEE Transactions on Wireless Communications}, 2025.

\bibitem{haupt2010antenna}
R.~L. Haupt, \emph{Antenna arrays: a computational approach}.\hskip 1em plus
  0.5em minus 0.4em\relax John Wiley \& Sons, 2010.

\bibitem{3gpp-tr-38-827}
\BIBentryALTinterwordspacing
{3GPP}, ``Study on radiated metrics and test methodology for the verification
  of multi-antenna reception performance of {NR} user equipment ({UE}),'' 3rd
  Generation Partnership Project (3GPP), Technical Report TR 38.827, 3GPP
  Technical Report. [Online]. Available:
  \url{https://portal.3gpp.org/ChangeRequests.aspx?q=1&specnumber=38.827}
\BIBentrySTDinterwordspacing

\end{thebibliography}
\end{document}